%% file: iclr2025_conference.tex
\documentclass{article} 
\usepackage{iclr2025_conference,times}

\input{math_commands.tex}

\usepackage{hyperref}       
\usepackage{url}            
\usepackage{booktabs}       
\usepackage{colortbl}
\usepackage{amsfonts}       
\usepackage{nicefrac}       
\usepackage{microtype}      
\usepackage{xcolor}         
\usepackage{microtype}
\usepackage{enumerate}
\usepackage{graphicx}
\usepackage{diagbox}
\usepackage{textcomp}
\usepackage{color}
\usepackage{tablefootnote}
\usepackage{xspace}
\usepackage{mathtools}
\usepackage{comment}
\usepackage{wrapfig}
\usepackage{caption}
\newcommand{\etal}{\textit{et al}.\ }
\usepackage{mathrsfs}
\usepackage{multirow}
\usepackage{algorithm}
\usepackage{algorithmicx}
\usepackage{algpseudocode}

\newcommand{\ouralg}{ProDiF\xspace}

\newcommand{\wsp}{\mathbf{W}_p}

\definecolor{mygray}{gray}{0.6}

\title{\ouralg: \underline{Pro}tecting \underline{D}omain-\underline{i}nvariant \underline{F}eatures  to Secure Pre-Trained Models Against Extraction}

\author{Tong Zhou$^{\dagger}$, Shijin Duan$^{\dagger}$, Gaowen Liu$^{\ddagger}$, Charles Fleming$^{\ddagger}$, Ramana Rao Kompella$^{\ddagger}$,   \\ 
[-1em]
    \AND \raisebox{1em}{Shaolei Ren$^{\mathparagraph}$ and Xiaolin Xu$^{\dagger}$}\\  [-1em]
$^\dagger$Northeastern University, $^\ddagger$Cisco, $^\mathparagraph$UC Riverside\\
\texttt{\{zhou.tong1, 
duan.s, x.xu\}@northeastern.edu, shaolei@ucr.edu,} \\ \texttt{\{gaoliu, chflemin, rkompell\}@cisco.com}
}

\iclrfinalcopy 
\begin{document}

\maketitle

\begin{abstract}
Pre-trained models are valuable intellectual property, capturing both domain-specific and domain-invariant features within their weight spaces. However, model extraction attacks threaten these assets by enabling unauthorized source-domain inference and facilitating cross-domain transfer via the exploitation of domain-invariant features. In this work, we introduce \ouralg, a novel framework that leverages targeted weight space manipulation to secure pre-trained models against extraction attacks.
\ouralg quantifies the transferability of filters and perturbs the weights of critical filters in unsecured memory, while preserving actual critical weights in a Trusted Execution Environment (TEE) for authorized users. A bi-level optimization further ensures resilience against adaptive fine-tuning attacks. Experimental results show that \ouralg reduces source-domain accuracy to near-random levels and decreases cross-domain transferability by 74.65\%, providing robust protection for pre-trained models. This work offers comprehensive protection for pre-trained DNN models and highlights the potential of weight space manipulation as a novel approach to model security. The code is available at: \url{https://github.com/Tongzhou0101/ProDiF}.
\end{abstract}

\section{Introduction}

Pre-trained deep neural networks (DNNs) excel across diverse tasks and encapsulate rich information in their weight spaces, making them valuable intellectual property (IP) \cite{xue2021dnn}. However, this richness also makes them vulnerable to model extraction attacks \cite{sun2021mind,dubey2022high}, enabling attackers to perform \textbf{source-domain inference} using the extracted models.
To counter this, prior works use Trusted Execution Environments (TEEs) to protect critical model weights, ensuring attackers obtain incomplete parameters, degrading source-domain performance \cite{chakraborty2020hardware,zhou2023nnsplitter}. However, these works overlook a critical post-leakage attack objective: \textbf{cross-domain transfer}.

The motivation behind attackers engaging in cross-domain transfer on victim models lies in the practical advantage of transferring the extracted model to other domains of interest. This exploitation allows attackers to leverage the \textit{domain-invariant features} (i.e., features that transfer across domains \cite{muandet2013domain}) acquired by the pre-trained model, overcoming data scarcity and saving effort compared to training models from scratch \cite{zhuang2020comprehensive}.  If domain-invariant features persist in the extracted model, attackers can achieve competitive cross-domain transfer, violating model owners' rights. While similar malicious transfers have been studied in MLaaS scenarios \cite{wang2022non,wang2023model,ding2024probe}, they do not address on-device pre-trained models or source-domain inference.

Considering both source-domain inference and cross-domain transfer as practical attack objectives for on-device ML following model extraction, we are motivated to propose a comprehensive protection method that can process pre-trained models to achieve \textbf{(i) accuracy degradation on the source domain } and \textbf{(ii) reduced transferability to target domains} for attackers.
To maintain high performance for authorized users, we also leverage TEE to secure critical model weights, allowing their participation in inference exclusively for authorized users.
However, implementing such solutions is intricate due to the following challenges: 1) How to define the possible target domain? 2) How to decide the critical weights? 3) How to ensure the resilience of the defense approach against adaptive attacks (e.g., attackers modifying the extracted model to improve accuracy)?



In this work, we propose \textit{\ouralg} to provide dual protection for pre-trained DNN models, addressing the three challenges outlined earlier. 
For the first challenge, while the specific target domain cannot be predicted, we assume its distribution is likely close to the source domain, a common premise in cross-domain transfer attacks \cite{torrey2010transfer}. Inspired by single-domain generalization \cite{qiao2020learning,volpi2018generalizing}, we generate auxiliary domains based on the source domain to address target domain uncertainty.
For the second challenge, we quantify transferability to identify filters learning domain-invariant features. To avoid channel mismatch and detection by attackers, we retain the full model in unsecured memory but perturb the weights of selected filters. These weights capture shared features between the source and potential target domains; perturbing them degrades performance on both domains. The perturbations are stored in TEE secure memory to maintain accuracy for authorized users.
Finally, we formulate a bi-level optimization problem to ensure the protected model's resilience against fine-tuning attacks by adversaries.

By applying \ouralg, we can leverage the rich resources in unsecured memory without concern for damage due to model extraction attacks.  Meanwhile, we utilize secure memory to store the benign weights of selected filters, preserving domain-invariant features for authorized users while preventing extraction by attackers.
The contributions of this work are summarized as follows:
\begin{itemize}
    \item To our knowledge, this is the first work that mitigates the consequences of model extraction by addressing expanding threats, including unauthorized source-domain inference and cross-domain transfer, thus achieving comprehensive protection for pre-trained DNN models. 
    \item Leveraging the generated auxiliary domains, \ouralg employs a bi-level optimization to optimize critical weights of the victim model, ensuring the resilience of the resulting protected model against adaptive fine-tuning attacks. 
    \item Experiments demonstrate that \ouralg achieves dual protection: it exhibits nearly random guessing performance on the source domain and reduces transferability by  74.65\% on potential target domains. Our results illustrate that obfuscating weights capturing domain-invariant features can substantially enhance model security.
   
\end{itemize}

\section{Background}
\subsection{Threat Model}
\label{sec:threat}

\begin{figure*}[t]
    \centering
    \includegraphics[width=0.95\textwidth]{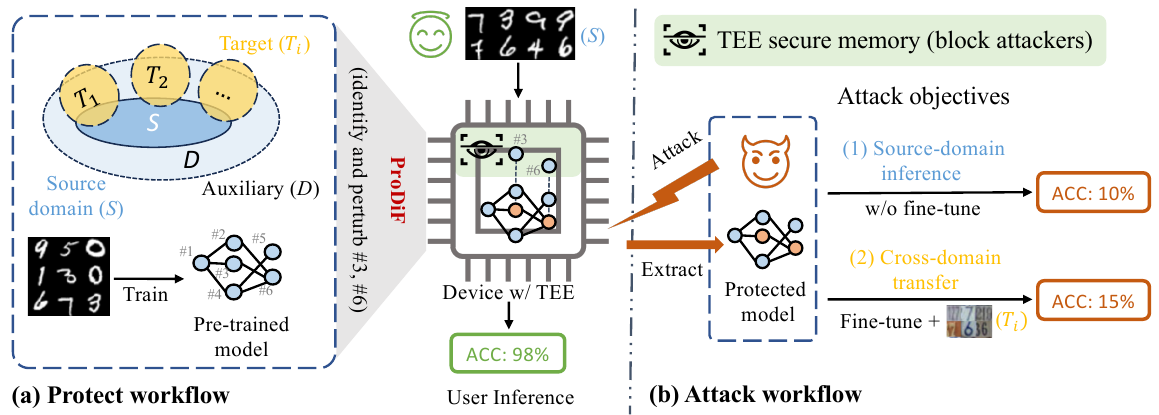}
    \caption{The workflow of protection and attack.
    For the model trained on the source domain, \ouralg identifies critical filters (e.g., \#3 and \#6) leveraging auxiliary domains, then employs bi-level optimization to perturb these weights (red dots) to generate the protected model. This model, stored in unsecured memory, effectively prevents attackers from source-domain inference and cross-domain transfer. The weight perturbations of critical filters stored in TEE secure memory can correct the features influenced by perturbed weights, enabling users to attain high performance.}
    \label{fig:overview}
\end{figure*}

\textbf{Attackers.} We assume strong attackers possess knowledge of the source domain of the victim model.  We also assume they have the capability to extract the on-device DNN model \textit{from unsecured memory}, including its architecture and pre-trained weights, e.g., using various side-channel attacks \cite{sun2021mind,dubey2022high}. Notably, this scenario differs from query-based model stealing attacks in MLaaS \cite{tramer2016stealing}, which use queried input-output pairs to train a function-equivalent model. 
As illustrated in Fig. \ref{fig:overview} (b), the attackers could exploit the extracted model in two ways:
\textbf{1)} they directly use the model for source-domain inference when their interests align with the source domain; \textbf{2)} they fine-tune the extracted model with \textit{a small set of collected data from the target domain} to achieve cross-domain transfer, with the model architecture remaining unchanged.

\textbf{Model IP owners.} To protect pre-trained models against illicit model exploitation while preserving the model performance for authorized users, model owners may employ defense solutions with hardware support (e.g., TEE), which can block access to critical data stored in secure memory. 
Specifically, model owners deploy the pre-trained model with perturbed critical weights in unsecured memory, while storing the benign values of critical weights in secure memory, as shown in Fig. \ref{fig:overview} (a). During inference, users can access this data, but extraction is prevented, functioning akin to a black box due to the properties of TEE \cite{chakrabarti2020trusted}.

It is worth noting that our intention is not to block \textit{all} potential target domains, as this would be impractical given the vast array of possibilities. Instead, the focus is on a common and practical scenario where target domains are proximate to the source domain and share the same label space.

\subsection{Trusted Execution Environment}
\label{sec:tee}
TEE is widely adopted across various hardware platforms, including edge devices~\cite{TrustZonefor}, CPUs~\cite{cen2017trusted}, and GPUs~\cite{nvidia}.  Such a hardware feature has already been widely used for DNN model protection by attributing to the  
hardware-based isolation between the secure and normal memory of a device to isolate sensitive information \cite{shadownet,chakraborty2020hardware,zhou2023nnsplitter,liu2023mirrornet,liu2024tbnet}. 
However, its secure memory space is relatively limited, e.g., 3-5MB in TrustZone \cite{amacher2019performance}, compared to the size of normal DNNs. 
Additionally, it is noted that even with authorized inference, the data inside the TEE remains a black box for authorized users. While they can leverage the data within the TEE to achieve high inference accuracy on the source domain, they cannot extract the data (i.e., critical model weights in our case).

\subsection{Limitation of Prior Works}\label{sec:related_work}
As model IP protection is critical for model owners, several studies have explored methods to degrade the model's performance to mitigate the risk of model extraction. For example, Chakraborty \etal propose a key-dependent training algorithm that obfuscates the weight space, causing a significant drop in accuracy when the model is extracted and used on different devices  \cite{chakraborty2020hardware}.
Moreover, Zhou \etal introduce a protection approach that obfuscates a small set of model weights using a fine-grained mask designed by a reinforcement learning-based controller \cite{zhou2023nnsplitter}. However, these methods primarily focus on degrading accuracy on the source domain, which cannot naturally reduce the transferability to prevent cross-domain transfer by attackers. 

Several works have proposed methods to restrict model transferability\cite{wang2022non,wang2023model,zhou2024archlock,ding2024probe}, where they ensure high performance in the source domain due to the focus on MLaaS, and thus cannot prevent unauthorized source-domain inference. Wang \etal propose a training strategy to restrict model transferability by capturing exclusive data representation in the source domain \cite{wang2022non}. Furthermore, the compact un-transferable isolation domain is proposed to prevent cross-domain transfers by emphasizing the private style features of the authorized domain \cite{wang2023model}. Also, \cite{zhou2024archlock} proposed a framework that restricts transferability from the architecture design perspective via neural network search, while \cite{ding2024probe} proposed a method designed for pre-trained encoders to prevent encoder generating meaningful embedding for unauthorized domains.
Importantly, ensuring comprehensive model protection is not as simple as combining the aforementioned methods, each of which is designed to address a specific vulnerability. These methods employ different training strategies, making seamless integration challenging \cite{wang2022non,wang2023model,chakraborty2020hardware}. Consequently, model owners cannot sequentially apply different methods to train the model and expect it to concurrently address both vulnerabilities.

\section{Proposed Method: \ouralg}\label{sec:method}

Aiming at achieving the best for preventing both illegal usages, i.e., source-domain inference and cross-domain transfer, we propose a comprehensive model IP protection approach \ouralg. 
It involves the following steps. First, auxiliary domains are generated to effectively address the inherent uncertainties within the target domain (Sec. \ref{sec:auxi}). Subsequently, critical weights are identified by ranking the transferability of convolutional filters, where filters exhibiting high transferability acquire domain-invariant features (Sec. \ref{sec:mask}). Finally, the protected model is derived by perturbing these critical weights through a bi-level optimization process (Sec. \ref{sec:bi-level}). The perturbed weights act as a barrier against attackers seeking to exploit the victim model, effectively safeguarding model IP. 

\subsection{Auxiliary Domains}\label{sec:auxi}

To address the uncertainty of target domains, we introduce a generative domain augmentation method to generate auxiliary domains ${\mathcal{D}_i}$ based on the source domain. 
The rationale behind this approach is that attackers are more likely to conduct model transfer when the target domain closely aligns with the source. This choice provides them with the advantages of transfer learning while avoiding potential drawbacks, as discussed in \cite{wang2019characterizing}. Therefore, while we cannot predict the specific target domain that attackers may select, we can approximate its distribution by augmenting the source domain.

Inspired by \cite{qiao2020learning}, we build a Wasserstein Auto-Encoder (WAE) parameterized by $\theta$ to introduce domain discrepancy between the source domain $\mathcal{S}$ and the auxiliary domains $\mathcal{D}_i$.  However, to simulate cross-domain transfer, we need to generate labeled samples, simulating how attackers use limited labeled data to fine-tune the extracted model. To this end, we enhance the WAE by incorporating the label as a condition in the decoding phase. 
Specifically, it consists of an encoder $Q(x|z)$ and a decoder $G(z,y)$, where $x$ and $y$ represent the input and its label, and $z$ represents the bottleneck embedding. The implementation details can be found in Appendix \ref{app:WAE}. Additionally, we employ Maximum Mean Discrepancy (MMD) to quantify the distance between $Q(x)$ and the Gaussian prior $P(z)$\cite{tolstikhin2018wasserstein}. The conditional WAE is optimized by:
\begin{equation}
  \min_{\theta}||G(Q(x),y))-x||_2 + \textnormal{MMD}(Q(x),P(z)).
  \label{eq:wae}
\end{equation}
After optimization, we can generate samples to construct auxiliary domains via the decoder $G$. To encourage a range of domain discrepancies, we introduce feature augmentation by incorporating a latent vector sampled from $\mathcal{N}~(\mu,\sigma)$ into the feature embedding. This augmentation technique has been proven effective in enhancing domain discrepancy through variations in $\mu$ and $\sigma$ \cite{qiao2021uncertainty}. As a result, we can generate diverse auxiliary domains $\mathcal{D}_i$. 
Each auxiliary domain is divided into a training set and a validation set for subsequent bi-level optimization.

\subsection{Filters Selection} 
\label{sec:mask}
The filter selection strategy aims to identify the filter with the highest transferability in each convolutional layer. This selection is based on the understanding that filters with higher transferability capture invariant patterns across domains, making them more intrinsically transferable than others \cite{phung2021learning}. 
In quantifying filter transferability, we adopt the approach similar to one in \cite{wang2019transferable}. We first measure the discrepancy $d$ between the source domain $\mathcal{S}$ and each auxiliary domain $\mathcal{D}_i$ for each channel $c$ (output of a filter) following Eq. \ref{eq:distance}, where we omit $i$ for simplicity.
\begin{equation}
  d^{(c)} = \left | \frac{\mu_{\mathcal{S}}^{(c)}}{\sqrt{{\sigma_{\mathcal{S}}^2}^{(c)} } } - \frac{\mu_{\mathcal{D}}^{(c)}}{\sqrt{{\sigma_{\mathcal{D}}^2}^{(c)} } } \right | , \label{eq:distance}
\end{equation}
where $\mu^{(c)} = \frac{1}{b}\sum_{j=1}^{b}\boldsymbol{x}^{(c)}_j$, 
${\sigma^2}^{(c)} = \frac{1}{b}\sum_{j=1}^{b}(\boldsymbol{x}^{(c)}_j-\mu^{(c)})$.
Here, $b$ denotes the batch size, and $\boldsymbol{x}$ represents the feature map of convolutional layers. We choose $\boldsymbol{x}$ from source domain when calculating $\mu_{\mathcal{S}}$ and ${\sigma_{\mathcal{S}}^2}$, and from $\mathcal{D}_i$ otherwise.
Hence, the distance-based probability $\boldsymbol{\alpha}$ for each channel c is calculated as:
\begin{equation}
\boldsymbol{\alpha}^{(c)} = \frac{K(1+d^{(c)})^{-1}}{\sum_{n=1}^{K}(1+d^{(n)})^{-1} }, \quad c= 1,2,...,K
\label{eq:alpha}
\end{equation}
where $K$ is the channel size, and a higher $\boldsymbol{\alpha}$ score indicates greater transferability of a filter.
We calculate the $\boldsymbol{\alpha}$ scores for each source-auxiliary domain pair and take the average $\boldsymbol{\alpha}$ to rank the filters.

Considering the limitations of secure memory in TEE, \ouralg selects only the filter with the highest transferability in each layer and stores its benign weights in secure memory. Next, we create a filter-wise binary mask, with 1 indicating the selected filters. This mask, denoted as $\mathbf{M}$, is applied to pre-trained model weights, allowing for the perturbation of weights exclusively for selected filters.
By adjusting the weights of these high-transferability filters via the following bi-level optimization, we disrupt the invariant features learned by both the source and potential target domains, thus achieving concurrent accuracy degradation and transferability reduction.

\subsection{Bi-level Optimization} 
\label{sec:bi-level}
We consider a pre-trained model $\mathcal{M}$ with weights $\mathbf{W}$ as the victim model, which is trained on labeled data pairs ($\boldsymbol{x_s}$, $\boldsymbol{y_s}$) from the source domain $\mathcal{S}$. 
To generate its protected version deployed in the unsecured memory, we perturb the weights corresponding to the chosen filters to $\wsp'$ (:=$\wsp$+$\Delta \wsp$) while keeping the rest fixed as $\mathbf{W}\setminus\wsp$.
If one only aims to achieve basic protection, i.e., preventing attackers from source-domain inference,
the optimization of modified weights $\wsp'$ follows:
\begin{equation}
    \max _{\wsp'}\mathcal{L} \left(f\left(\boldsymbol{x_s} ;\wsp' \cup (\mathbf{W}\setminus\wsp) \right), \boldsymbol{y_s}\right),
    \label{eq:basic}
\end{equation}
where $f$ denotes the functionality of the DNN, and $\mathcal{L}$ denotes the loss function of the source task. 

Considering that our protective measures extend beyond safeguarding source-domain inference to actively preventing cross-domain transfer, we enhance the aforementioned method by introducing a bi-level optimization approach to achieve our objective. More importantly, we aim to find a perturbation that enhances the resilience of the protected model. In doing so, we aspire to ensure that even if potential attackers manage to fine-tune the extracted model using a limited set of target domain data, they can still only achieve poor performance.

\textbf{Lower-level Optimization}. In this stage, we emulate how attackers optimize the extracted model for transfer by finding an optimal weight update $\Delta \mathbf{W}$ on the extracted model to enhance performance in the target domain. This mimicking is done using training data ($\boldsymbol{x_i^{tr}}$, $\boldsymbol{y_i^{tr}}$) in the auxiliary domain $\mathcal{D}_i$. We optimize $\Delta \mathbf{W}$ with the following equation:
\begin{equation}
    \Delta\mathbf{W}^*_{i} = \arg\min _{\Delta \mathbf{W}_{i}}\mathcal{L} \left(f\left(\boldsymbol{x_i^{tr}} ;\mathbf{W}' + \Delta\mathbf{W}_{i}\right), \boldsymbol{y_i^{tr}}\right)
    \label{eq:lower}
\end{equation}
where $\mathbf{W}'$ is the weights of the protected model obtained from the upper-level optimization.

\textbf{Upper-level Optimization}. To leverage the knowledge gained from the lower-level optimization regarding the transferability of the model with current weights $\mathbf{W}'$, we proceed to optimize the weights $\mathbf{W}'$ further. The goal here is to diminish the transferability of the model across all auxiliary domains. This process is guided by the following equation:
{\small
\begin{align}
    \max  _{\wsp'}&\sum_{i=1}^{\mathcal{I}} \lambda_i\mathcal{L} \left(f\left(\boldsymbol{x_i^{val}} ;\mathbf{W}'+\Delta\mathbf{W}^*_{i}\right), \boldsymbol{y_i^{val}}\right), \label{eq:upper} \\
    \text{s.t.} \; \mathbf{W}' = & \wsp'\cup(\mathbf{W}\setminus\wsp) \notag 
    = \wsp' \odot \mathbf{M} + \mathbf{W}  \odot (\mathbf{1}-\mathbf{M}) \notag
\end{align}
}

where $\lambda_i$ is the scaling factor of the loss on ($\boldsymbol{x_i^{val}}, \boldsymbol{y_i^{val}}$), i.e., the samples and labels from the validation set in the auxiliary domain $\mathcal{D}_i$
and $\sum_{i=1}^{\mathcal{I}} \lambda_i = 1$. Notably, we also incorporate the source domain into the auxiliary domain during the bi-level optimization to ensure accuracy degradation on the source domain.

Integrating all the aforementioned strategies, we present \ouralg in Algorithm \ref{alg:overall}, with the details of the interaction with TEE shown in Appendix~\ref{app:summary}.

\section{Experiments}
\subsection{Experiment Setup}
\label{sec:settings}
\textbf{Datasets and models.} 
We examine our methods on three types of datasets as previous works \cite{wang2022non,wang2023model}. One is the \textbf{digits datasets}, including MNIST (MN) \cite{lecun1998gradient}, USPS (US) \cite{hull1994database}, and
SVHN (SV) \cite{netzer2011reading}. 
The training and test sets for these digit datasets are unbalanced, i.e., the number of each class is unequal.
Another is the \textbf{natural image datasets}, including CIFAR10  \cite{krizhevsky2009learning} and STL10 \cite{coates2011analysis}. The last one is VisDA \cite{peng2017visda}, which is a large-scale dataset, including over 280K \textit{synthetic} (training set) and \textit{real} (validation set) images across 12 categories.
The details of the dataset and data processing are shown in Appendix \ref{app:dataset}.
Following \cite{wang2022non,wang2023model}, we use VGG-11 \cite{simonyan2014very} for digits classification,  VGG-13 \cite{simonyan2014very} for natural image datasets, and ResNet-50 \cite{he2016deep} for VisDA.

\textbf{Settings.}
We generate auxiliary domains based on the source domain, where $\mu$ is from 0 to 1 with a step of 0.25 and $\sigma$ is set to 0.5 and 1. The examples are shown in Appendix \ref{app:example} and the hyperparameter settings can be found in Appendix \ref{app:hyper}.
To emulate the cross-domain transfer process using a limited dataset that potential attackers may possess, we employ 5\% of the training data from the target domain for simulation.
Note that the target domain remains unseen during \ouralg protection. 
Also, for \ouralg protection, we only select the filter with the highest transferability in each convolutional layer, excluding the one with a 1$\times$1 kernel. The ratio of selected critical weights to the total parameters for each model remains unchanged, where the required storage is small enough ($<$ 133KB) to meet the secure memory constraint (3-5MB) on TEE (see Appendix \ref{app:tee}).
Moreover, we use top-1 inference accuracy (\%) as the evaluation metric for all experiments, which are implemented using PyTorch and conducted on NVIDIA A10 GPU.



\subsection{Comparison Methods}
As \ouralg is the first approach to simultaneously prevent source-domain inference and cross-domain transfer (i.e., dual protection), there are no previous works fully comparable to it.
However, to better evaluate the effectiveness of \ouralg, we adopt the following SOTA works as a comparison on a single aspect. For protecting source-domain inference, 
we compare with the SOTA method \textit{NNSplitter \cite{zhou2023nnsplitter}}, which shares the same objective as the basic protection mentioned in our work. As for defending against model transfer, the recent works \textit{NTL \cite{wang2022non}} and \textit{CUTI \cite{wang2023model}} can be used for comparison. 
Moreover, we also propose \textit{Baseline}, where no protection is applied to the victim model, and  \textit{Supervised Learning}, which simulates the scenario that attackers initialize the extracted model and train it from scratch with limited data (5\% of training data).

\begin{table*}[t]
\renewcommand\tabcolsep{7pt}
\centering
\caption{The results of defending against source-domain inference. The lower accuracy (\%) indicates better protection on the source domain. The baseline is the benign accuracy of victim models.  
}
\label{tab:res_direct}
\resizebox{.9\linewidth}{!}{
\begin{tabular}{ccccccc}
\toprule[0.4mm]
                             & \multicolumn{3}{c}{\textbf{Digits}} & \multicolumn{2}{c}{\textbf{Natural}} & \multicolumn{1}{c}{\textbf{VisDA}} \\
                                \cmidrule(r){2-4} \cmidrule(r){5-6}
                                \cmidrule(r){7-7}
                                & MNIST       & USPS      & SVHN      & CIFAR10            & STL10           & Synthetic                    \\ \midrule
Baseline               &   99.03          &   95.61        &     93.57      &    94.22                &     82.19            &         93.92                          \\
Supervised Learning         &79.18            &      82.96       &    68.35                  &      46.21              &     19.11            &         49.16                          \\ 
\textit{NTL} \cite{wang2022non}         & 97.9         &    98.8         &     88.4                 &     88.9               &      86.2                               &       92.4        \\
\textit{CUTI}  \cite{wang2023model}        &99.1         &     99.6        &     90.9                 &   83.9                 &     86.8            &                   94.1               \\ 
\textit{NNSplitter} \cite{zhou2023nnsplitter}        & 9.90            &     12.31      &   8.87        &        10.00            &    10.00                                 &      9.41         \\ \midrule
\textbf{\ouralg} &  10.04$\pm$0.29           &   13.25$\pm$1.22        &    8.90$\pm$2.45       &           10.00$\pm$0.00         &   10.00$\pm$0.00               &   8.07$\pm$1.74           \\\bottomrule[0.4mm]
\end{tabular}}
\end{table*}
\subsection{Performance Evaluation}
\label{sec:eval}
Based on our threat model, we consider two attack scenarios and demonstrate that \ouralg can defend against them at the same time, while SOTA works can only achieve one aspect.

\textbf{Scenario I: Source-domain Inference.} When attackers' target domains align with the source domain of the victim model, they can directly utilize the extracted model to conduct inference without further fine-tuning. 
By intentionally introducing perturbations to the model weights, \ouralg ensures that the extracted model performs poorly and becomes ineffective for direct use by attackers. 
From the results in Tab.~\ref{tab:res_direct}, we can observe that by applying \ouralg, the top-1 inference accuracy experiences a significant drop compared to the baseline (i.e., from over 90\% to $\sim$10\% on digits), where the baseline applies no protection to the victim model, allowing attackers to achieve the exact inference accuracy as the victim model. 
Besides, the success of \ouralg in accuracy drops results from misclassifying all images into a random class. Specially, since all classes in CIFAR10 and STL10 are balanced, the accuracy is degraded to 10\% with no variance.

Moreover, it is noted that \ouralg can achieve a comparable degraded accuracy as the SOTA work \textit{NNSplitter}. The degraded accuracy is even lower than the victim DNN trained with supervised learning, rendering the source-domain inference useless. However, \textit{NTL}  and \textit{CUTI}, which were specifically designed to restrict model transferability, fall short in preventing source-domain inference, as evidenced by their high accuracy on the source domain.

\begin{table*}[t]
\centering
\caption{Accuracy in cross-domain transfer is denoted by $\Rightarrow$, with the source domain on the left and the unseen target domain on the right. The average drop (Avg. Drop) measures the deviation from the baseline, where a greater drop indicates better protection.
}
\label{tab:res}
\resizebox{\linewidth}{!}{
\begin{tabular}{ccccccc|c}
\toprule[0.4mm]
      & MN $\Rightarrow$ US & MN $\Rightarrow$ SV & US $\Rightarrow$ MN & US $\Rightarrow$ SV & SV $\Rightarrow$ MN  & SV $\Rightarrow$ US & Avg. Drop \\ \midrule
 Baseline & 93.35   &  81.53  &  93.20  &  81.97  &  94.78  &  94.67 & /\\
 Supervised Learning & 82.96 & 68.35 & 79.18 & 68.35 & 79.18 & 82.96 & 13.09\\
\textit{NTL} \cite{wang2022non}   &  13.8  &  20.8  &  6.7  & 6.0   &  12.3  &  8.9 & \textbf{78.50}\\
\textit{CUTI} \cite{wang2023model}   &  6.7  &  6.7  &  9.1  & 25.5   &  11.9  &  14.3 & \textbf{77.55}\\
\textit{NNSplitter} \cite{zhou2023nnsplitter}   &  92.67  &  80.52  &  93.11  & 81.71   &  94.34  &  94.40 & 0.46\\
\textbf{\ouralg}  &  17.88$\pm$0.23  &  19.58$\pm$0.47  &  11.35$\pm$0.49  &  19.59$\pm$0.38  & 10.10$\pm$0.76   &  13.15$\pm$0.39 & \textbf{74.65}\\ \bottomrule[0.4mm]
                          
\end{tabular}}
\end{table*}


\begin{wrapfigure}{r}{0.55\textwidth} 
    \centering   
    \vspace{-5pt}
    \includegraphics[width=0.54\textwidth]{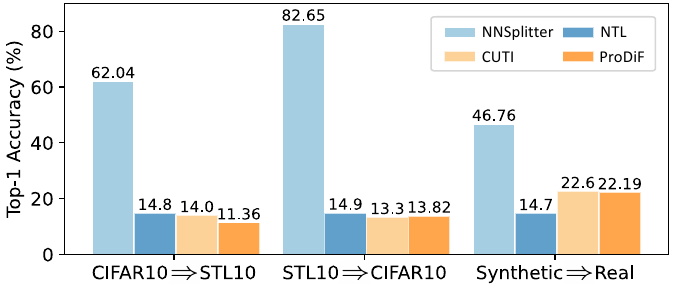}
    \caption{The results of CIFAR10, STL10, and VisDA. The lower accuracy indicates better protection against cross-domain transfer.}
    \label{fig:transfer}
\end{wrapfigure}
\textbf{Scenario II: Cross-domain Transfer.}  
We form several domain pairs to demonstrate that the models under \ouralg protection have reduced transferability, where the target domain is unknown during protection. 
As indicated by the results on digits in Tab.~\ref{tab:res}, while \textit{NNSplitter} exhibits optimal performance in defending against source-domain inference, it demonstrates almost no reduction (only a 0.46\% accuracy drop) in transferability compared to the baseline. 
These results highlight that accuracy reduction on the source domain does not necessarily impact model transferability.
In contrast, \ouralg proves effective in reducing transferability, e.g., 84.68\% accuracy drop for SV $\Rightarrow$ MN. 
Furthermore, it attains protection comparable to \textit{NTL} and \textit{CUTI}, with a difference of less than 4\% in transferability reduction.
For the transfer results on other domain pairs, depicted in Fig.~\ref{fig:transfer}, 
\ouralg outperforms \textit{NNSplitter} and achieves a protection performance similar to \textit{NTL} and \textit{CUTI}. Importantly, unlike \textit{NTL} and \textit{CUTI}, \ouralg does not require input-level modifications, such as adding a special patch, for transferability reduction.

Overall, \ouralg achieves comprehensive model protection by defending against source-domain inference and model transfer at the same time, filling the gap of SOTA methods. It successfully reduces the accuracy of the source domain to the random-guessing level (e.g., $\sim$10\% for ten-class classification tasks), while restraining the performance on unseen target domains lower than attackers training one from scratch using supervised learning.

\begin{figure*}[t]
    \centering
    \includegraphics[width=\textwidth]{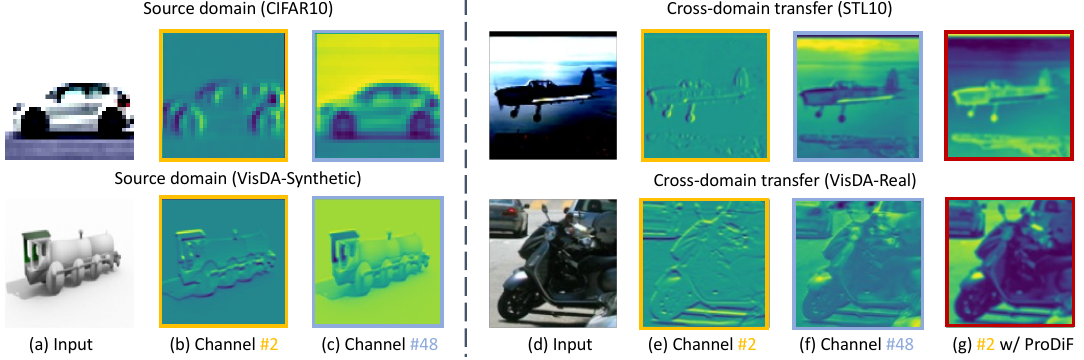}
    \caption{High-transferability filters mainly extract edge information (orange box, channel \#2), whereas low-transferability filters capture more diverse details (blue box, channel \#48). \textbf{With \ouralg protection (red box), the domain-invariant feature will be perturbed from (e) to (g). }}
    \label{fig:feature}
\end{figure*}
\subsection{Filter Selection} 
The filter selection strategy in the mask design of \ouralg is based on the transferability measurement of each channel (Sec.~\ref{sec:mask}). By selecting the critical weights of filters associated with the highest transferability, we identify the domain-invariant features, e.g., \textbf{the one learns the edge of the object} (as shown in Fig.\ref{fig:feature} with an orange box). To protect domain-invariant features from attackers, \ouralg perturbs these weights to disrupt the domain-invariant features. As shown in Fig.\ref{fig:feature} (g), \textit{after perturbation, the filter learning edge now just learns the varying details as the one with the lowest transferability.} More examples are shown in Appendix \ref{app:feature}. We also include a random filter selection strategy to demonstrate the effectiveness of our selection strategy in Appendix~\ref{app:random}.


\subsection{Optimization Method}
To illustrate its effectiveness, we compare it with a \textit{naive} model protection scheme employed by the protector. In this scenario, we assume the protector possesses knowledge of the target domain akin to the attacker and has access to a small dataset from that domain. The protector first simulates the attacker's optimization approach by  enhancing the extracted model accuracy and then optimizes again to deliberately degrade accuracy for protection, following an alternative optimization procedure outlined in Eq.~(\ref{eq:att}) and Eq.~(\ref{eq:pro}):
\begin{equation}
\small
    \text{(attack) }\Delta\mathbf{W}^*_{a} = \arg\min _{\Delta \mathbf{W}_{a}}\mathcal{L} \left(f\left(\boldsymbol{x_{t}} ;\mathbf{W}'+\Delta\mathbf{W}_{a}\right), \boldsymbol{y_{t}}\right)
    \label{eq:att}
\end{equation}
\begin{equation}
\small
\text{(defense) }\max _{\wsp'} \mathcal{L}\left(f\left(\boldsymbol{x_{t}} ;\wsp'\cup\left(\mathbf{W}\setminus\wsp)+\Delta\mathbf{W}^*_{a} \right)\right), \boldsymbol{y_{t}}\right)
\label{eq:pro}
\end{equation}

where 
$\Delta\mathbf{W}^*_{a}$ is the weights modification by attackers via transfer learning,  $\wsp'$ is optimization from the protector, and ($\boldsymbol{x_{t}},\boldsymbol{y_{t}}$) denoted 5\% labeled data from the target domain. Unlike bi-level optimization in \ouralg, here both attackers and protectors use specified target datasets, and optimize the same model one after the other.
Here we set the size of labeled data ($\boldsymbol{x_{att}},\boldsymbol{y_{att}}$) in the target domain to 1,000, and the settings of hyper-parameters are kept the same as in the previous experiments.

\begin{wraptable}{r}
{0.6\textwidth} 
\centering
\vspace{-12pt}
\caption{Comparison of transferability reduction using different optimization methods.}
\label{tab:naive}

\resizebox{\linewidth}{!}{
\begin{tabular}{cccc|c}
\toprule
 & MN$\Rightarrow$SV & US$\Rightarrow$MN & SV$\Rightarrow$US & \textbf{Avg. Drop} 
 \\ \midrule
Naive  &    79.58$\pm$0.95      &  90.88$\pm$1.17       & 92.46$\pm$0.47 &  2.16 \\
  Bi-level    &  19.58$\pm$0.47                        &  11.35$\pm$0.49     &        10.10$\pm$0.76 & \textbf{76.12} 
\\
\bottomrule
\end{tabular}}
\end{wraptable}

The model performance after 50-step naive optimization is reported in Tab.~\ref{tab:naive}. Compared to the baseline method, the naive optimization only shows  2.16\% accuracy drop, while the bi-level optimization achieves over 70\% accuracy drop. 
The reason behind this difference lies in the nature of the naive optimization process. As described in Appendix \ref{app:loss}, the naive optimization is dynamic, meaning that due to the same set of samples ($\boldsymbol{x_{t}},\boldsymbol{y_{t}}$) but with opposite objectives in the two equations, each equation can counteract the modifications made by the other in the last step. 
As the attacker is the one who makes the final modification on the extracted model, they can always improve the accuracy to a level comparable to the baseline. This explains the limited effectiveness of the naive optimization approach in reducing model accuracy compared to the bi-level optimization approach.

\section{Conclusion}
In this work, we introduce \ouralg, a novel framework for protecting on-device DNN models against source-domain inference and cross-domain transfer while preserving model performance for users using TEE. 
By selectively modifying a small subset of weights in the pre-trained model, \ouralg achieves near-random guess performance on the source domain and transferability reduction for potential target domains. The original values of these modified weights are stored in TEE, which is only accessible to authorized users. Moreover, \ouralg remains effective regardless of how attackers fine-tune the protected model. Unlike prior works that focus on single types of attacks, ours tackles the complexities of dual threats, filling a gap in existing defenses that often fail when confronted with diverse attack scenarios. Overall, \ouralg stands out for its comprehensive approach to combating expanding attack surfaces.

\section{Acknowledgment}
This work is supported in part by the U.S. NSF under Grants CNS-2326597, CNS-2239672,  and a Cisco Research Award.
\bibliography{ref}
\bibliographystyle{iclr2025_conference}

\newpage
\appendix

\section{Implementation of WAE}
\label{app:WAE}
The design of the encoder $Q$ and the decoder $G$ follows the prior work \cite{tolstikhin2018wasserstein}.  The input size of $Q$ is $32\times32\times1$ for digit datasets and $32\times32\times3$ for CIFAR10 \& STL10 \& VisDA. Specifically, the encoder architecture is (using input size of $32\times32\times3$ as an example): 
\begin{align*}
x\in\mathbb{R}^{32 \times 32 \times 3} &\to \mathrm{Conv}_{128} \to \mathrm{BN} \to \mathrm{ReLU}\\
&\to \mathrm{Conv}_{256} \to \mathrm{BN} \to \mathrm{ReLU}\\
&\to \mathrm{Conv}_{512} \to \mathrm{BN} \to \mathrm{ReLU}\\
&\to \mathrm{Conv}_{1024} \to \mathrm{BN} \to \mathrm{ReLU} \to \mathrm{FC}_{64},
\end{align*}
and the decoder architecture is:
\begin{align*}
z \in \mathbb{R}^{74} &\to \mathrm{FC}_{4 \times 4 \times 1024}\\
&\to \mathrm{ConvTranspose}_{512} \to \mathrm{BN} \to \mathrm{ReLU}\\
&\to \mathrm{ConvTranspose}_{256} \to \mathrm{BN} \to \mathrm{ReLU}\\
&\to \mathrm{ConvTranspose}_{128} \to \mathrm{BN} \to \mathrm{ReLU} \to \mathrm{FSConv}_{3},
\end{align*}
where ConvTranspose denotes transposed convolutional layers. Here the input of decoder is the concatenation of the output latent vector of $Q$ (64-dimension) and the one-hot encoding of the 10-class labels (12 classes for VisDA).

Both architectures include 4 convolutional layers with a filter size of 5$\times$5. The stride is set to 2 for the first 3 Conv and ConvTranspose in $Q$ and $D$, respectively.
The WAE is optimized for 100 epochs according to Eq. \ref{eq:wae} using the Adam optimizer. The learning rate is initially set to 0.001 and decreased by a factor of 2 after 20 epochs.

\section{Summary of \ouralg}
\label{app:summary}

\begin{algorithm}[htbp]
\small
\caption{\ouralg}
\label{alg:overall}
\textbf{Input}: pre-trained model $\mathcal{M}$; auxiliary domains $\mathcal{D}$$_i$ \\
\textbf{Output}: protected model $\mathbf{W}'$
\begin{algorithmic}[1] 
\Statex\textcolor{mygray}{// Generate mask}
\State Calculate $\boldsymbol{\alpha}^{(c)}$ across $\mathcal{D}$$_i$  \Comment{Eq. (\ref{eq:alpha})}
\State Generate a filter-wise mask $\mathbf{M}$ 
\Statex\textcolor{mygray}{// Apply basic protection as a warm-up}
\State Initialize $\wsp'$ with $\mathbf{M}$  constrained \Comment{Eq. (\ref{eq:basic})}
\Statex\textcolor{mygray}{// Bi-level optimization loop}
\Repeat \label{alg:loop_start}
\State $\mathbf{W}'\leftarrow \wsp'\cup (\mathbf{W}\setminus\wsp)$
\For{all $\mathcal{D}$$_i$}
\State Obtain $\Delta\mathbf{W}^*_i$ in lower-level stage \Comment{Eq. (\ref{eq:lower})}
\State Calculate loss with  $\mathbf{W}'$+$\Delta\mathbf{W}^*_i$ 
\EndFor
\State Update $\wsp'$ in upper-level stage \Comment{Eq. (\ref{eq:upper})}
\Until{Converge} \label{alg:loop_end}
\end{algorithmic}
\end{algorithm}

The algorithm has the following key points:
(i) Identify the most critical filters and generate a corresponding mask $\mathbf{M}$; 
(ii) Initialize  $\wsp'$ by applying basic protection; (iii) Employ a bi-level optimization to further optimize $\wsp'$. 
Here we also add $\mathcal{S}$ as one $\mathcal{D}_i$ during bi-level optimization to ensure that accuracy degradation is maintained on the source domain.

Besides, the perturbation of benign weights $\Delta\wsp$ are stored in TEE secure memory, which corrects the functionality of the protected model for authorized users. 
In particular, for the perturbed channel in a specific layer, we compute the convolution of weights $\wsp'$ and input features ($X_i$) in the unsecured memory. This yields an output feature map ($O_n$) containing errors in that output channel. However, for users with access to secure memory, the convolution of input features with $\Delta\wsp$ is computed. This result (denoted as $O_s$) will be subtracted from $O_n$ to correct errors in specific output channels, thereby obtaining benign features for normal users (i.e., $X_{i+1} := \wsp * X_i = O_n-O_s $, by virtue of the distributive nature of the convolution operation). Therefore, authorized users will obtain the high accuracy as the high-performing victim model with no accuracy loss.

\section{Experiments}
We evaluate the performance of \ouralg on multiple DNN models and datasets for the classification tasks. Unlike prior methods focusing solely on source-domain inference or cross-domain transfer, we will demonstrate that \ouralg achieves dual protection, excelling in both aspects with performance comparable to state-of-the-art (SOTA) methods.

\section{Details of Datasets}
\label{app:dataset}

\textbf{Digits.} \textit{MNIST} contains 70k gray-scale images of handwritten digits from 0 to 9, each of which is 28$\times$28 pixels in size. The training set includes 60k labeled data and the rest data are for testing. \textit{USPS} has 7291 training and 2007 test images in gray-scale with size 16$\times$16. It was created by taking a sample of real-world handwritten digits from envelopes and intended to be representative of the variation in handwritten digits that might be encountered in the wild. 
\textit{SVHN} is a widely used digits dataset and more challenging for classification tasks than \textit{MN}, which consists of real-world images collected from  house number plates. It contains over 70k 32$\times$32 RGB images for training and over 20k for testing. The training set and test set for all these digit datasets are unbalanced, i.e., the number of each class is unequal, shown in the Tab.~\ref{tab:digit}.

\vskip 0.1in
\begin{table}[htbp]
\centering
\caption{The number of each class in the digits dataset.}
\label{tab:digit}

\resizebox{0.7\linewidth}{!}{
\begin{tabular}{cccccccccccc}
\toprule
         & & 0 & 1 & 2 &3 &4&5&6&7&8&9 \\ \midrule
  \multirow{3}{*}{Training}&MNIST    & 5923& 6742& 5958& 6131& 5842& 5421&5918&6265& 5851& 5949   \\ 
 & USPS    & 1194& 1005&731&658& 652& 556& 664& 645& 542& 644  \\
  & SVHN    & 4948& 13861& 10585& 8497& 7458& 6882& 5727&5595& 5045& 4659   \\
\midrule

\multirow{3}{*}{Test} & MNIST    & 980& 1135& 1032& 1010& 982& 892&958&1028& 974& 1009    \\ 
 & USPS    & 359& 264&198&166& 200& 160& 170& 147& 166& 177   \\
  & SVHN    & 1744& 5099& 4149& 2882& 2523& 2384& 1977&2019 &1660& 1595   \\
   
\bottomrule
\end{tabular}}
\end{table}

\begin{figure}[htbp]
    \centering
    \includegraphics[scale=0.9]{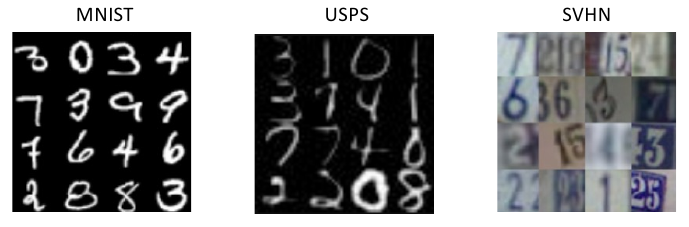}
    \caption{The visualization of the digits datasets. }
    \label{fig:digit}
\end{figure}

We also apply the same data processing to all digits datasets, i.e., resizing all images to 32$\times$32. We also apply the gray-scale transform to images in SVHN to be consistent with MNIST and USPS. As the visualization of each digit dataset shown in the Fig.~\ref{fig:digit}, the SVHN dataset is more complex than the the MNIST and USPS datasets, making the classification task more challenging. Therefore, the model trained on SVHN will learn more complicated feature representation and can be transferred to the MNIST/ USPS domain with high accuracy. 

\textbf{Natural Image}. Both CIFAR-10 and STL-10 have 10 classes. For CIFAR-10, it contains 60k 32$\times$32 color images in 10 classes, with 6k labeled images per class. It is divided into 50k  training images and 10k test images. As for STL-10, it has 5k training images and 8k test images with the original size of 96$\times$96. To make the DNN models applicable to both datasets, we down-scale the images to 32×32 resolution to match that of CIFAR-10.

\textbf{VisDA}. We experiment with two domains in VisDA classification dataset, i.e., synthetic renderings of 3D models, and the real-image domain. Each domain contains 12 classes. The details of each domain are shown in Tab. \ref{tab:visda}. Following \cite{wang2022non}, we resize the images to 112$\times$112 for both domains.

\begin{table}
    \centering
    \caption{The number of each class in two domains of the VisDA dataset. }
    \label{tab:visda}
    \resizebox{\linewidth}{!}{
    \begin{tabular}{ccccccccccccc}
    \toprule
       Class  &  Plane&  Bicycle&  Bus&  Car&  Horse&  Knife&  Motorcycle&  Person& Plant & Skateboard&Train &Truck\\ \midrule
         Synthetic&  14309&  7365&  16640&  12800&  9512&  14240&  17360&  12160&  10371& 11680& 16000&9600\\
         Real&  3646&  3475&  4690&  10401&  4691&  2075&  5796&  4000&  4549& 2281& 4236&5584\\ \bottomrule
    \end{tabular}}
    
\end{table}




\section{Hyper-parameters}
\label{app:hyper}
After attackers extract the pre-trained model, they will either directly use it in the same domain, or apply transfer learning with 1000 data in the target domain to improve the accuracy by fine-tuning the whole model. To mimic the real-world behavior of both attacker and protector, 
we construct a small searching space with common choices [0.01, 0.005, 0.001, 0.0005. 0.0001] for each side, and finally determine the initial learning rate for the protector is 0.0005, and for the attacker is 0.001. The learning rate will then decay at a rate of 0.5 every 10 epochs. 

For the fine-tuning training strategy, we follow the same setting in Sec. 5.1, and the training process stops when the inference accuracy does not improve for the last 3 validation epochs.

\section{Examples of auxiliary domains}
\label{app:example}
\begin{figure} [h]
    \centering
    \includegraphics[width=0.9\textwidth]{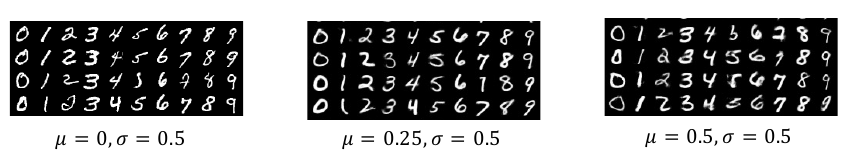}
    \caption{The examples of auxiliary domains for MNIST.}
    \label{fig:aux}
\end{figure}

\begin{figure} [t]
    \centering
    \includegraphics[width=0.75\textwidth]{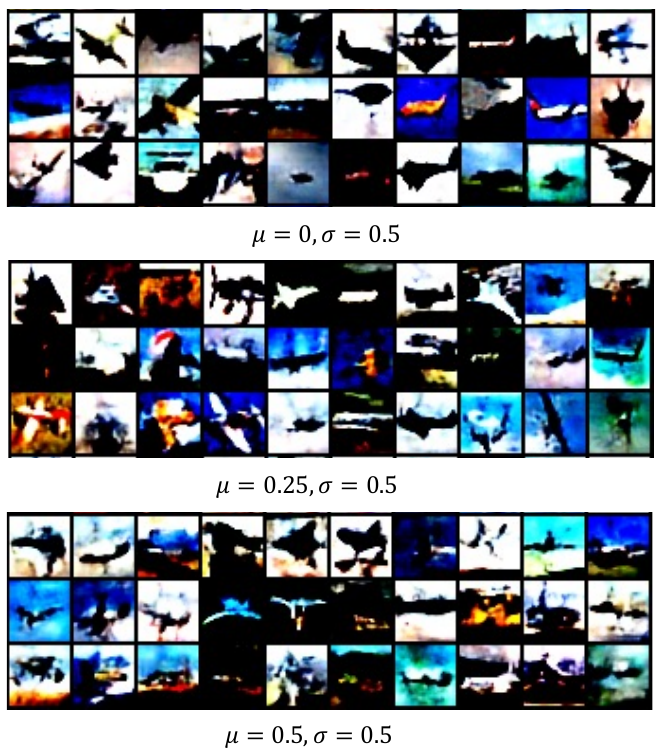}
    \caption{The examples of auxiliary domains for CIFAR10.}
    \label{fig:aux_cf}
\end{figure}

We provide a few examples of generated auxiliary domains in Fig. \ref{fig:aux} and Fig. \ref{fig:aux_cf}. We also measure the MMD between the source domain and its auxiliary domains, and we observe the MMD increases with the value of $\mu$ when $\sigma$ is fixed. For example, in Fig. \ref{fig:aux}, the MMD is [0.017, 0.036, 0.041] when $\mu$ =[0, 0.25, 0.5] and $\sigma$ = 0.5.

\begin{figure} [h]
    \centering  
    \includegraphics[scale=0.4]{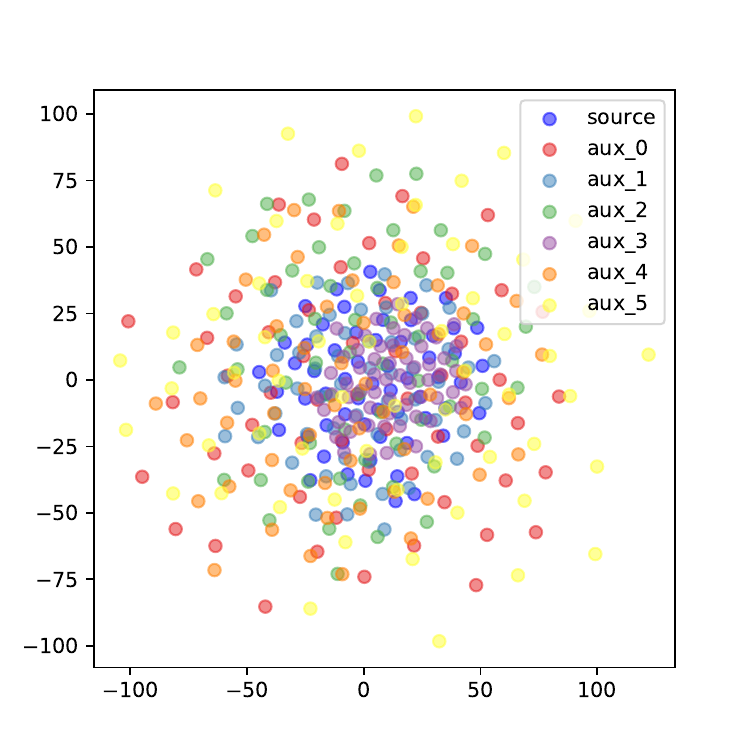}
    \caption{The visualization of MNIST and its auxiliary domains using t-SNE}
    \label{fig:tsne}
\end{figure}
Besides, we use t-SNE to visualize the latent space of the source domain and the auxiliary domains, as shown in Fig. \ref{fig:tsne}. Here we sample 100 data from 6 auxiliary domains for better visualization, where $\mu$=[0, 0.5, 1] and $\sigma$=[0.5, 1]. The Fig. \ref{fig:tsne} successfully demonstrates that the auxiliary domains can augment the source domain.

\section{TEE}
\label{app:tee}
Since the secure memory in TEE is stringent, we aim to identify a subset of critical model weights that affect the model performance for both the source domain and the target domains. In our experiment, the ratio of selected critical weights to the total parameters for each model remains is shown in Tab.~\ref{tab:tee}. Besides, in the extreme case that the model is very large, we can change the current filter selection strategy to a more fine-grained weight selection, i.e., only selecting important weights in these filters, making the TEE secure memory compatible for applying \ouralg to such large models.
\begin{table}[h]
    \centering
     \caption{The statistics of critical weights in each victim model.}
    \label{tab:tee}
    \resizebox{.4\linewidth}{!}{
    \begin{tabular}{cccc} 
    \toprule
         &  \# Total para.&   Ratio& Storage\\ \midrule
         VGG-11& 28.13M & 0.07\% & 79KB  \\ 
         VGG-13& 28.33M & 0.08\% &  87KB \\ 
         ResNet-50& 23.52M &  0.15\%  & 133KB\\ \bottomrule
    \end{tabular}}   
\end{table}

Without authorized access to the critical weights stored in TEE, the extracted model part will perform poorly due to a lack of critical data. 
Since the effectiveness of TEE has been previously established in other studies \cite{chen2019deepattest,shadownet}, this work will not delve into the implementation details.

\section{Analysis of Naive Optimization}
\label{app:loss}
In the application of naive optimization can lead to a back-and-forth process between the protector and the attacker. Initially, the protector modifies critical weights in the pre-trained model to degrade its accuracy in the target domain. Subsequently, the attacker extracts the model and performs fine-tuning to improve its accuracy. This iterative exchange resembles a strategic online game, where each side responds to the actions of the other.

\begin{figure}[htbp]
    \centering   \includegraphics[width=0.95\linewidth]{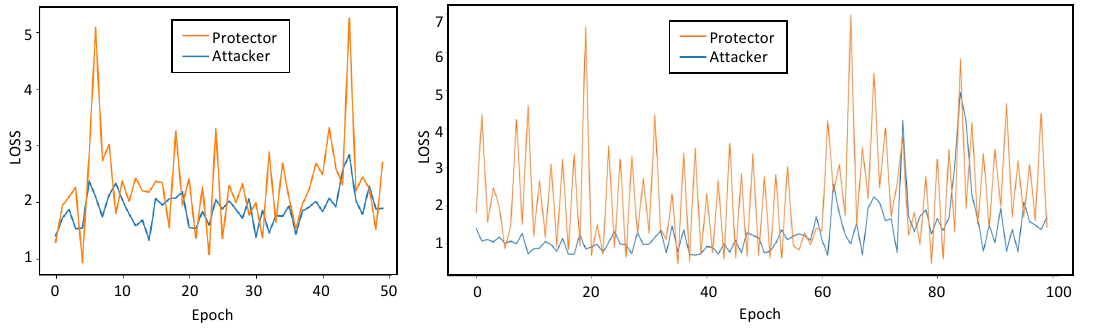}
    \caption{The loss of naive optimization. }
    \label{fig:loss}
\end{figure}

The resulting effect is depicted in Figure \ref{fig:loss}, where both sides experience drastic and non-converging changes in their losses, even when extending the training epoch from 50 to 100. In contrast, the bi-level optimization approach implicitly considers the attacker's actions, strategically thinking ahead by anticipating future steps, leading to improved performance.

\vskip 0.1in

\section{Random Filter Selection}
\label{app:random}

\begin{figure}[h]
    \centering
    \includegraphics[scale=0.7]{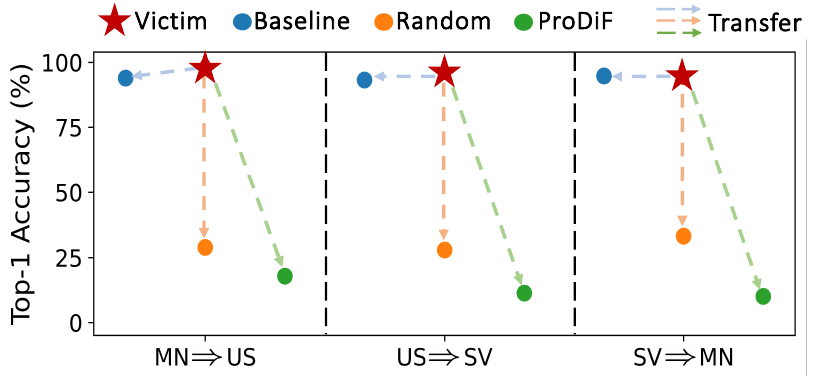}
    \caption{The comparison of random filter selection filter selection with \ouralg.  The dot marks show the accuracy of the victim model after being transferred to the target domain. }
    \label{fig:random}
\end{figure}

To measure the influence of our filter selection strategy based on ranking transferability, we replace it with a random filter selection strategy, i.e., randomly selecting filters to perturb their weights for model protection. We conduct the comparison experiments on digits datasets, and the results are shown in Fig.~\ref{fig:random}, where the performance of the victim is measured on the source domain. Overall, random filter selection can also reduce transferability compared to the baseline among all cases, but \ouralg performs even better with an extra 14.95\% accuracy drop on average. 

\section{Feature Visualization}
\label{app:feature}
\begin{figure}[htbp]
    \centering
    \includegraphics[width=0.8\linewidth]{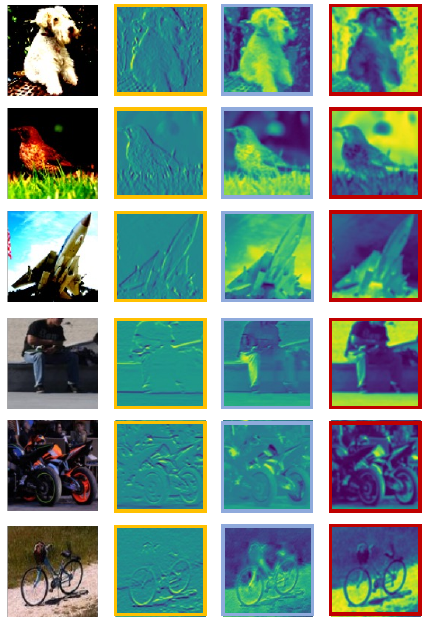}
    \caption{Comparison of domain-invariant (orange box) and domain-specific (blue box) features. Orange box features are generated from filters with high transferability, while blue box features are from filters with low transferability. Results of disrupted orange box features are shown in the red box.}
    \label{fig:stl}
\end{figure}

The success of \ouralg can be attributed to the optimization of critical weights achieved through the bi-level optimization process. The critical weights correspond to the filters that generate domain-invariant features, i.e., features in orange boxes as shown in Fig.~\ref{fig:stl}. The domain-invariant features learn more sharing patterns like textures and edges of objects. In contrast, the filters with the lowest transferability learn domain-specific features like varying backgrounds.

\section{Modification and Fine-tuning Range}
\label{sec:range}
\begin{wraptable}{r}
{0.48\textwidth} 
\centering
\caption{The comparison of different modification and fine-tuning ranges, where the source domain is MN and the target is US.}
\label{tab:range_direct}
\resizebox{0.95\linewidth}{!}{
\begin{tabular}{cccc}
\toprule
 & Protector & Attacker   & \ouralg\\ \midrule
 \multirow{2}{*}{\begin{tabular}[c]{@{}c@{}}w/o fine-tune\\ (Source)\end{tabular}} &p1 & - &      12.59$\pm$3.60 \\
& p2 & -                              &  10.04$\pm$0.29
\\\midrule
\multirow{4}{*}{\begin{tabular}[c]{@{}c@{}}w/ fine-tune\\ (Target)\end{tabular}} & p1 & a1 &  58.07$\pm$0.43\\
& p1 & a2 &  39.36$\pm$0.44\\
&p2 & a1 &  22.27$\pm$0.33\\
&  p2 & a2 &  17.88$\pm$0.23\\

\bottomrule
\end{tabular}}

\end{wraptable}
We explore different modification ranges for protectors and model fine-tuning ranges for attackers. For protectors, the choices include modifying: p1) the first layer, and p2) one filter per layer. For attackers, they can either conduct source-domain inference on the source domain without fine-tuning or fine-tuning a1) the last layer or a2) the whole extracted model for transferring to the target domain. We evaluate the performance of each modification and fine-tuning range on the VGG-11 model trained on MN.
The results in Tab.~\ref{tab:range_direct} show that p2 outperforms p1 when defending against the source-domain inference. 
When adopting p2, attackers can only achieve low accuracy ($\sim$20\%) no matter if they fine-tune the last layer or the whole model. While the results are limited to the scope that attackers use a specific optimizer for fine-tuning, it still demonstrates the advantage of spreading the weight perturbation across all layers. Also, similar to~\cite{deng2024sophon}, we can also apply meta-learning to optimizers to ensure the protection performance even if attackers use a different one.


\newpage

\end{document}

%% file: math_commands.tex
\usepackage{amsmath,amsfonts,bm}

\def\eqref#1{equation~\ref{#1}}

\def\1{\bm{1}}

\DeclareMathAlphabet{\mathsfit}{\encodingdefault}{\sfdefault}{m}{sl}
\SetMathAlphabet{\mathsfit}{bold}{\encodingdefault}{\sfdefault}{bx}{n}

